\documentstyle[sprocl]{article}

\def\Journal#1#2#3#4{{#1} {\bf #2}, #3 (#4)}

\def\NPA{{\em Nucl. Phys.} A}

\def\PLB{{\em Phys. Lett.}  B}
\def\PRL{\em Phys. Rev. Lett.}
\def\PLC{\em Phys. Rep.}

\def\PRD{{\em Phys. Rev.} D}
\def\PR{\em Phys. Rev. }

\begin{document}

\title{ COLOR SUPERFLUIDITY AND CHIRAL SYMMETRY BREAKDOWN IN DENSE QCD MATTER}

\author{ JI\v{R}\'{I} HO\v{S}EK}

\address{  Institute of Nuclear Physics,
\v{R}e\v{z} near Prague, CZ 250 68, Czech Republic}


\maketitle\abstracts{We describe the interplay of two nonperturbative phenomena
which should take place in the chirally invariant deconfined phase of QCD
matter at finite density and $T$=0 : (i) Cooper-pair quark-quark ground-state
condensation in appropriate channels should yield exotic sorts of color
superfluidity, and (ii) quark-antiquark ground-state condensation should yield
spontaneous breakdown of chiral symmetry. We briefly review the main recent
achievements in the subject, and present a field-theory formalism which enables
to deal with both above mentioned types of condensates simultaneously. }

\section*{1.Introduction}

Long ago  a great colleague of ours theoretically speculated that
``by convention there is color, by convention sweetness, by
convention bitterness, but in reality there are atoms and space.''
(Democritus, 400 B.C.) Rather ironically, what Democritus called
conventions we call reality today. At this conference it is
appropriate to paraphrase him as :``By convention there are atoms,
by convention mesons, by convention light nuclei, but in reality
there are colors, flavors and space'':
\begin{equation}
{\cal L}_{\rm QCD} =
-\frac{1}{4}F_{a\mu\nu}F^{\mu\nu}_{a}+\sum_{\rm F=1}^{\rm n_{F}}
\overline{\psi}_{\rm F}(i\!\not\!\!D -m_{\rm F}+ \mu_{\rm F}
\gamma_{0})\psi_{\rm F} \, . \label{lqcd}
\end{equation}
Starting from (\ref{lqcd}) I will discuss some properties of a system of two
massless $(m_{\rm F} = 0)$ up and down flavors $({\rm n_{F}} =2)$ of the
colored quarks (color indices of quarks are suppressed in (\ref{lqcd})) in the
deconfined phase characterized by large baryon number density, and zero
temperature. I will call the light flavors conservatively up and down although
it would be more elegant to call them sweet and bitter. By construction the
subject belongs to nuclear physics, and it became rather hot recently.

At high baryon number densities, for example in cores of the neutron stars,
hadrons overlap. Due to asymptotic freedom the QCD matter should behave as a
weakly interacting quark-gluon plasma \cite{col}. Restricting ourselves to
temperatures low enough that the thermal wavelength becomes comparable with
interparticle spacing, we can expect behavior of the system characteristic of a
genuine quantum Fermi liquid.

What are the interactions which determine the behavior of the deconfined
colored quark and gluon excitations in this phase ? First of all, there are no
experimental data, either real or the lattice ones which would provide check of
our considerations. From the theoretical point of view we argue as follows :
First, the electric components of gluons acquire perturbatively a mass due to
Debye screening \cite{col}. This effect gives rise to an effective fourquark
interaction of the form
\begin{equation}
{\cal L}^{\rm (e)}= G_{\rm e}
(\overline{\psi}\gamma_{0}\frac{1}{2}\lambda_{a}\psi)^{2}  \, ,
\end{equation}
 where $G_{\rm e}^{-1/2}\sim 300$MeV. Second, the magnetic components of gluons can be
considered as weak external perturbations and neglected in the lowest
approximation. They become massive nonperturbatively in the next step due to
the dynamical Higgs mechanism. This effect gives rise to an effective fourquark
interaction
\begin{equation}
{\cal L}^{\rm (m)}=G_{\rm m}(\overline{\psi}\gamma_{i}\frac{1}{2}
\lambda_{a}\psi)^{2} \, ,
\end{equation}
where $G_{\rm m}\sim G_{\rm e}$. Third, at not very large densities there is an
effective fourquark interaction originating from quark couplings with the
instanton zero modes \cite{hoo,diak1} :
\begin{equation}
{\cal L}^{\rm (i)} = G_{\rm i}[(\overline{\psi}\psi)^{2}+
(\overline{\psi}i\gamma_{5}\vec{\tau}\psi)^{2}-
(\overline{\psi}\vec{\tau}\psi)^{2}-(\overline{\psi}i\gamma_{5}\psi)^{2}]
\, ,
\label{inst}
\end{equation}
again with $G_{\rm i} \sim G_{\rm e}$. This interaction has the important and
desirable property of breaking explicitly the global axial U(1) symmetry.
Fourth, there are no gluons at zero temperature. Fifth, any massive
dynamically generated collective excitation gives rise to a local fourquark
interaction the exact form of which cannot be determined a priori. It follows
that the dynamics of QCD matter in the deconfined phase at low $T$ is governed
by an effective local fourquark ${\rm SU(2)_L \otimes SU(2)_R}$ chirally
invariant Lagrangian in which the color ${\rm SU(3)_{c}}$ symmetry is treated
as global :
\begin{equation}
{\cal L}_{\rm eff}=\overline{\psi}( i\!\not\!\partial+
\mu\gamma_{0})\psi+ {\cal L}^{\rm (e)}+ {\cal L}^{\rm (m)}+ { \cal
L}^{\rm (i)}+ \dots
\label{leff}
\end{equation}
On the first sight all these interactions seem to be irrelevant in Wilson's
sense when scaling the momenta towards Fermi surface. One is inclined to expect
that the system will indeed behave as a (relativistic) Landau Fermi liquid. The
argument contains a potential loophole \cite{pol} : If any of the considered
interactions contains an attraction between two quarks at the Fermi surface
with momenta $\vec{p}_{\rm F}$ and $-\vec{p}_{\rm F}$, respectively, then this
very component becomes marginal, and leads to a phase transition. The effect is
well known in nonrelativistic Fermi systems as Cooper instability. It is
responsible for the BCS superconductivity, and for the superfluidity of
${}^3$He .

We are then faced with an important question : Do the quark interactions in the
deconfined low-$T$ phase contain ``dangerous'' quark-quark attractive channels
? The answer is yes, even in the naive perturbative regime : The one-gluon
exchange is attractive not only in the color-singlet quark-antiquark channel
$(3 \otimes \overline{3}= 1 \oplus 8)$, but with a half strength also in the
color-antitriplet quark-quark channel $(3 \otimes 3=\overline{3} \oplus 6)$.
Hence, we should expect a quark Cooper-pair condensation in attractive channels
and, consequently some sort of color superfluidity or superconductivity. With
perturbative forces in mind the idea was mentioned already in \cite{col}, and
elaborated in \cite{frau,bai}. Recent interest
\cite{alf1,rap,ber,car,alf2,step} is concentrated mainly on the interaction
(\ref{inst}).

\section*{2. Playing with condensates}

With an appropriate  local fourfermion  interaction between quarks
carrying spin, isospin and color there can be just four types of
superfluid condensates different from zero due to Pauli principle
:
\begin{eqnarray}
v_{(1)} & =&\langle\overline{\psi}A\tau_{2}\gamma_{5}\psi^{\cal C}
\rangle \, ,\label{v1} \\
v_{(2)} &=&
\langle\overline{\psi}A\tau_{3}\tau_{2}\gamma_{0}\gamma_{3}\psi^{\cal C}\rangle
\, , \label{v2}\\
v_{(3)}&=&\langle\overline{\psi}S\tau_{3}\tau_{2}\gamma_{5}\psi^{\cal C}\rangle
\, , \label{v3} \\
 v_{(4)}&=&\langle\overline{\psi}S\tau_{2}\gamma_{0}\gamma_{3}\psi^{\cal C}\rangle
 \, . \label{v4}
\end{eqnarray}
Although we use the convenient Lorentz-covariant notation with
$\psi^{\cal{C}}=C\overline{\psi}$ where $C$ is the charge-conjugation matrix,
it is clear that the only sacred property of the ground state here is the
translational invariance. Consequently, we contemplate four distinct ordered
phases :
\begin{enumerate}
\item Condensate $v_{(1)}$ with the color-antisymmetric Clebsch-Gordan (CG) matrix
$A$ chosen as $A^{ab}=i\epsilon^{ab3}$ corresponds to a
ground-state expectation value of the order parameter $\Phi^c$. It
describes a Lorentz scalar, isosinglet, color-triplet superfluid.

\item Condensate $v_{(2)}$ corresponds to a ground-state
expectation value of the order parameter $\Phi^c_{I;\mu\nu}$. It describes a
color-triplet superfluid which at the same time behaves as ordinary, as well as
flavor, ferromagnet ($\Phi^{c}_{I;\mu\nu}$ is an isospin one, antisymmetric
tensor field).

\item Condensate $v_{(3)}$ with the color-symmetric CG matrix $S$
chosen as $S^{ab}=\frac{1}{3} \delta^{ab}- \frac{1}{\sqrt{3}} (\lambda_8)^{ab}$
corresponds to a ground-state expectation value of the order parameter
$\Phi^{{ab}}_{I}$. It describes a Lorentz scalar color-sextet superfluid which
at the same time behaves as a flavor ferromagnet.

\item Condensate $v_{(4)}$ corresponds to a ground-state
expectation value of the order parameter $\Phi^{{ab}}_{\mu\nu}$. It describes
an isoscalar color-sextet superfluid which at the same time behaves as an
ordinary ferromagnet.
\end{enumerate}

Recent papers are devoted predominantly to the first case with some estimates
made \cite{alf1} also of the fourth possibility.

It is important to realize that all terms in  (\ref{v1}-\ref{v4}) are invariant
with respect to global chiral SU(2) rotations. This implies that they cannot
produce the physical, chiral-symmetry-violating quark masses in the deconfined
phase. There is, however, a ground-state condensate which can do namely this,
\begin{equation}
\langle\overline{\psi}\psi\rangle \neq 0
\label{scal}
\end{equation}
and it should be considered together with (\ref{v1}-\ref{v4}). It is well
established  that the interactions (\ref{leff}) for the numerical values of the
couplings above critical and for densities below critical give rise to the
quark masses and, by virtue of the Goldstone theorem, also to the massless
pions. In fact, spontaneous breakdown of chiral symmetry at finite density
yields nontrivial subtleties revealed only recently \cite{alf1,bub}.

\section*{3. Self-consistent perturbation theory}

Quantitative formulation of the physical program described above is the
following : Given the effective ${\rm SU(3))_{c} \otimes SU(2)_{L} \otimes
SU(2)_{R}}$ invariant Lagrangian of the type (\ref{leff}) with arbitrary
couplings $G$, analyze its spectrum and properties of its deconfined ordered
phases characterized by the vacuum condensates (\ref{v1}-\ref{scal}) different
from zero. First real insight into the problem was provided by the  BCS-like
variational calculation \cite{alf1}. Here I present main steps of a
generalization of the field-theoretical approach developed for
superconductivity by Nambu \cite{nam1}, and for the spontaneous chiral symmetry
breaking in the nucleon-pion system by Nambu and Jona-Lasinio \cite{nam2}. It
gives almost immediately the dispersion laws of true fermionic quasiparticles
as exemplified explicitly below . It also provides a systematic way of
investigating the gapless collective Nambu-Goldstone excitations corresponding
to all symmetries spontaneously broken by the condensates
(\ref{v1}-\ref{scal}). This part of the program will be published separately.
For illustration, for comparison with the results of others, and in order to
keep the formulas relatively simple I present characteristic results assuming
that only $v_{(1)}$ and $\Sigma$ are different from zero and generated by the
interaction ${\cal L}^{\rm (i)}$. The general formulas will be published
separately.

First, the Lagrangian ${\cal L}_{\rm eff}=\overline{\psi}(i\!\not\!\partial+
\mu\gamma_{0})\psi +{\cal L}^{\rm (i)}$ is split as ${\cal L}_{\rm eff}= {\cal
L}_{0}^{'}+{\cal L}_{\rm int}^{'}$. The unperturbed ground state defined by
${\cal L}_{0}^{'}={\cal L}_{0}-
 {\cal L}_{\Delta}- {\cal L}_{\Sigma}$,
\begin{equation}
{\cal L}_{0}^{'}=\overline{\psi}(i\!\not\!\partial+
\mu\gamma_{0})\psi
-\frac{1}{2}[\overline{\psi}A\gamma_{5}\tau_{2}\Delta\psi^{\cal{C}}+H.c.]-
\overline{\psi}\Sigma\psi
\end{equation}
is supposed to be energetically favorable with respect to the one
defined by ${\cal L}_{0}=\overline{\psi}(i\!\not\!\partial+
\mu\gamma_{0})\psi$. Selfconsistency is achieved by imposing the
condition that the lowest-order perturbative contribution of
${\cal L}_{\rm int}^{'} = {\cal L}_{\rm int}+ {\cal L}_{\Delta}+
 {\cal L}_{\Sigma}$
 to  ${\cal L}_{0}^{'}$, using the propagator defined by ${\cal L}_{0}^{'}$,
 vanishes. This condition fixes the numerical values of $\Delta$ and $\Sigma$ in
terms of the dimensional parameters of the model. Trivial solutions
$\Delta=\Sigma=0$ always exist and correspond to the naive perturbative
expansion.

Second, the main trick consists of introducing  the field $q$
which is defined as follows:
\begin{equation}
q^{a}_{\alpha A}= \frac{1}{\sqrt{2}}\,
 \left(\begin{array}{c}
\psi^a_{\alpha A}(x)\\ [0.1in]
P^{ab}(\tau_2)_{AB}\delta_{\alpha\beta}\psi^{\cal C}_{\beta bB}(x)
\end{array}
\right)
\end{equation}
in which $P^{ab}\equiv [e^{i\alpha}A + e^{i\sigma}S]^{ab}$ has the property
$AS=0,\,  P^{+}P=1$. The field $q$ operates in space of Pauli matrices
abbreviated as $\Gamma_i$. In terms of $q$ the Lagrangian of the main interest
is
\begin{equation}
{\cal L}_0^{'}=\overline q \left[
\begin{array}{cccc}
\not\!p-\Sigma+\mu\gamma_0 & \; & \; & -\Delta\gamma_5\\[0.1in]
-\gamma_0(\Delta\gamma_5)^{+}\gamma_0 & \; & \; & \not\!p-\Sigma-\mu\gamma_0
\end{array}
\right]q\equiv \, \overline q S^{-1}(q)q.
\end{equation}
Third, in terms of $q$ the interaction Lagrangian (\ref{inst})
becomes
\begin{equation}
{\cal L}^{\rm (i)}=   G_{\rm i}[(\overline qq)^2+(\overline
q\Gamma_3i\gamma_5\vec{\tau}q)^2-(\overline
q\Gamma_3\vec{\tau}q)^2-(\overline qi\gamma_5q)^2].
\end{equation}
Fourth, the rest is a well defined manual work :

\begin{enumerate}
\item Explicit form of
\[
S(p) \equiv \left(
\begin{array}{cc} I & J \\ K & L
\end{array}
\right) \]
is the following :
\begin{eqnarray}
I & = & \frac{p_{+}+\Sigma}{D_{+}} \left[
1+\frac{|\Delta|^{2}}{D}(\not\!P+C)(\not\!p_{+}+\Sigma) \right] \, , \label{I}
\\
K &=&
\frac{\Delta^{*}}{D}\gamma_{5}(\not\!P+C)(\not\!p_{+}+\Sigma) \, ,
\label{K}
\end{eqnarray}
and analogous formulas hold for $L$ and $J$, respectively. Symbols
in (\ref{I}) and (\ref{K}) are defined as follows :
\begin{eqnarray}
p^{\mu}_{+} &\equiv &[(p_{0}+\mu),\vec{p}\, ] \, , \nonumber\\
D_{+}&\equiv &(p_{0}+\mu)^{2}-\epsilon_{p}^{2} \, , \ \, {\rm with} \
\epsilon^{2}_{p} = \vec{p}^{\, 2}+\Sigma^{2} \, , \nonumber\\
P^{\mu} & \equiv &
[D_{+}(p_{0}-\mu)-|\Delta|^{2}(p_{0}+\mu),(D_{+}-|\Delta|^{2})\vec{p}\, ] \, ,
\nonumber\\
C &\equiv& -(D_{+}-|\Delta|^{2})\Sigma \, , \nonumber\\
 D & \equiv & [(p_{0}+\mu)^{2}-\epsilon_{p}^{2}]
\nonumber\\
&& \left\{ p_{0}^{2}-[(\mu+\epsilon_{p})^{2}+|\Delta|^{2}] \right\}
\left\{p^{2}_{0}-[(\mu-\epsilon_{p})^{2}+|\Delta|^{2}]\right\} \, . \label{D}
\end{eqnarray}

\item The coupled nonlinear equations for $\Sigma$ and $\Delta$
have the form
\begin{eqnarray}
\Sigma&-& 13i G_{\rm i} \int \frac{d^4 p}{(2\pi)^4}\, tr(I+L)= 0 \, ,
\label{sigma}
\\
\Delta^{*} &-& 8i G_{\rm i}\int \frac{d^4 p}{(2\pi)^4}
(K\gamma_{5}+\gamma_{5}K) = 0 \, . \label{delta}
\end{eqnarray}
For further progress it is absolutely crucial that the denominator $D$ of the
propagator $S$, which defines the fermionic spectrum, has been found in the
factorized form : The $p_{0}$ integration can be performed explicitly using
Cauchy theorem, while the remaining integral over $d^{3}p$ can be either cut
off at some physical cutoff $\Lambda$, or regularized by the formfactor
$F(p^{2})=[\Lambda^{2}/(p^{2}+\Lambda^{2})]^{\nu}$ as suggested in \cite{alf1}.
In any case : if $\Delta \neq 0, \Sigma\neq 0$ are found by solving
(\ref{sigma}) and (\ref{delta}), then the lowest-order selfconsistent
perturbation theory turns the system of interacting massless quarks into a
system of noninteracting massive quasiquarks with the dispersion laws
explicitly given by (\ref{D}).

\item For $\Delta=0$ the equation (\ref{sigma}) is the standard gap equation
for the dynamical quark mass in the popular NJL approach \cite{nam2}. For
$\Sigma=0$ we have verified that the gap equation (4.4) of \cite{alf1} is
obtained in an approximation of keeping only the leading term of the strongest
residue in (\ref{delta}).

\item In the vacuum sector ($\mu=0$) all formulas greatly simplify,
and the form of our coupled vacuum gap equations
\begin{eqnarray}
\Sigma&- & 104i G_{\rm i} \int \frac{d^{4}p}{(2\pi)^{4}}
\frac{\Sigma}{p^{2}-\Sigma^{2}-|\Delta|^{2}} = 0 \, , \\
\Delta^{*} & - & 16i
G_{\rm i} \int \frac{d^{4}p}{(2\pi)^{4}}
\frac{\Delta^{*}}{p^{2}-\Sigma^{2}-|\Delta|^{2}} =0 \, ,
\end{eqnarray}
is in accord with the formulas of \cite{diak2}.
\end{enumerate}

\section*{4.Outlook}

Recent semiquantitative analyses of the deconfined quark
low-temperature matter using the realistic numerical values of the
parameters reveal the potential relevance of the ordered phases
for the heavy-ion collisions. In general, the vacuum engineering
with new peculiar macroscopic quantum phases is very promising,
and it is worth of being developed theoretically in all possible
details. Fortunately, this seems feasible. We find gratifying that
the NJL-like effective field-theory treatment is fully justified
for the description of the deconfined phase. Of course, it has to
be supplemented with a prescription which properly mimics the
asymptotic freedom of the underlying QCD.

\section*{Acknowledgments}
 This work  was supported by grant GA CR 2020506.

\end{document}